\documentclass[aps,amsfonts,showpacs,nofootinbib,
eqsecnum,floats,amssymb,amsmath,txfonts]{revtex4}
\usepackage{graphicx}
\usepackage{amsmath}
\usepackage{amsfonts}
\usepackage{amssymb}
\usepackage{pxfonts}
\usepackage{yfonts}
\def\@ApndToks#1#2{\edef\@act{\noexpand#1={\the#1#2}}\@act}

\def\@begineqtableau#1#2{
  \vcenter\bgroup\openup1\jot
    \mathsurround=0pt  \everymath={\displaystyle}
    \dimen0=#1  \count0=#2  \toks0={\strut}  \toks1={##}
        \def\\{\crcr\noalign{\vskip\dimen0}}
    \ifnum\count0 > 0
      \elloop  \advance\count0 by -1
        \@ApndToks{\toks0}{$\hfil\the\toks1$&${}\the\toks1\hfil$}
      \ifnum\count0 > 0
        \@ApndToks{\toks0}{&}
      \repeat
    \else  \@ApndToks{\toks0}{\hfil$\the\toks1$\hfil}  \fi
    \edef\@act{\noexpand\ialign\bgroup\the\toks0\noexpand\crcr}  \@act}
\def\@endeqtableau{\crcr\egroup\egroup}
\makeatother
\def\beq{\begin{eqnarray}}
\def\eeq{\end{eqnarray}}
\def\be{\begin{equation}}
\def\ee{\end{equation}}

\def\={\triangleq}
\def\nn{\nonumber\\}

\def\ub#1{\underleftarrow{#1}}

\def\mbf#1{\mbox{\boldmath${#1}$}}
\def\del{\partial}
\def\grad{\nabla}
\def\lie{\pounds}
\def\ellie{\pounds}

\def\bm{\bar m}
\def\bu{\bar U}
\def\bepsilon{\bar\epsilon}
\def\balpha{\bar\alpha}
\def\bbeta{\bar\beta}
\def\bpi{\bar\pi}
\def\bmu{\bar\mu}
\def\blambda{\bar\lambda}

\begin{document}
\title{Non- minimally Coupled Scalar Fields, Holst Action and Black Hole
Mechanics}

\author{Ayan Chatterjee}\email{ayan.chatterjee@saha.ac.in}

\affiliation{Theory Division, Saha Institute of
Nuclear Physics, Kolkata 700064, India}

%%%%%%%%%%%%%%%%%%%%%%%%%%%%%%%%%%%%%%%%%%%%%%%%%%%%%%%%%%%%%%%%%%%%%%
\begin{abstract}
The paper deals with the extension of the
Weak Isolated Horizon (WIH) formulation
to the non- minimally coupled scalar fields.
In the first part of the paper, we 
construct the appropriate Holst type action to incorporate non-minimal scalar
field and construct the covariant phase space of the theory.
Using this covariant phase space, we prove the laws of black hole
mechanics and show that with a gauge fixing, the symplectic structure on the
horizon reduces to that of a $U(1)$ Chern-Simons theory.
The level of the Chern- Simons theory is shown to depend
on the non-minimally coupled scalar field.
\end{abstract}

\pacs{04070B, 0420}

\maketitle

%%%%%%%%%%%%%%%%%%%%%%%%%%%%%%%%%%%%%%%%%%%%%%%%%%%%%%%%%%%%%%%%%%%%%%%%
\section{Introduction} 

General theory of relativity (GR) is the formost theory of gravity and has passed 
a number of experimental tests. However,
some future experiment might reveal
deviations from GR. This is expected because GR
can be thought of as an effective description
of a fundamental theory of underlying quantum structure.
Effective classical limits of quantum gravity
should contain all possible interactions with dimensionful length parameters
representing scales at which new degrees of freedom emerge to play crucial
roles. The idea is quite similar to the Fermi theory of weak
interactions. It is expected that as one goes higher in the energy scale,
newer degrees of freedom emerge. In other words,
the degrees of freedom relevant to quantum gravity are supressed
at lower energies (for example, the energy scale at Tevatron) and are
revealed only at scales of Planck length.
Thus, it is very natural to construct 
other effective theories which have
a limit to GR (at lower energies) while
show considerable deviations at higher energies.
Examples of such theories include the Brans-Dicke theory, the Einstein-Cartan
theory, extra dimensional models inspired by String theories and many more.

The scalar-tensor theories are parhaps the most popular alternative to
GR. The first of its kind was the Brans-Dicke theory \cite{bd}.
In conformity to the equivalence principle, this theory also
considers gravity in terms of spacetime curvature.
In addition, there exists a massless scalar field in the spacetime which
together with the gravitational constant $G$, determines the coupling
strength of gravity to matter. In some limit, the standard GR equations are
recovered. 
This theory is interesting because it
includes the possibility of variation of  effective gravitational
constant influenced by the scalar field, which can be constrained by 
direct astronomical observations like the solar system test.
In this paper, we shall not commit ourselves to a particular
form of the scalar-tensor theory
like the Brans-Dicke. Instead, the most general coupling
will be studied. It must be emphasised that these theories are different
from GR. To exemplify,
consider the Brans-Dicke 
theory with a non-minimal scalar coupling in the so called Jordan
frame (or String frame) with metric $g$.
In the Einstein frame with metric $\bar g$ (conformally related to $g$ by the
scalar field), the theory can be equivalently represented by a theory of a
minimally coupled scalar field in curved spacetime.
It might then seem that the theory is the same
as GR (with scalar field sources) in different variables. The crucial
difference is that test
particles move
along the geodesics determined by $g$ and will not (in general) coincide with
that of $\bar g$ \cite{bd}. One is free to choose the frame
for comfortable calculations. In the Jordan frame, the gravitational field
equations are different and 
in the Einstein frame, one has to account for the changes in the matter
equations. Either of these leads to deperture from GR.

The aim of the  present paper is to extend the formalism of Weak
Isolated Horizons (WIH) for non-minimally coupled scalar fields
(for details of isolated horizon formulation see \cite{abf} and for
its applications, see \cite{abl}).
The rationale for such an extension is the following.
We have argued that alternate effective
descriptions of gravity (other than GR) can exist in which new degrees of
freedom can become important at higher energies (say the Planck scale).
Black holes are ideal laboratories to look for
effects of such non-standard degrees of freedom. If a non-minimally coupled
scalar field becomes important at Planck scales, it can leave its imprint on the
entropy of these black holes. (String theories for example, predict
existence of such scalar fields.) Thus, if one is able to determine the
entropy of these black hole horizons, the exact dependence on these scalar
fields will be clarified. The \emph{loop approach} is one of the many ways of
calculating entropy \cite{abck,abk} (see \cite{mitra} for other approaches).
Here, one uses the formalism of isolated horizons to determine the effective
topological theory on the black hole horizon, quantise it and count the states.
Through a canonical analysis of the Palatini action, it was determined that this
topological theory is a $U(1)$ Chern-Simons theory \cite{ack}. Entropy is then 
calculated by quantising the Chern-Simons theory \cite{abck,abk}. 
Holst's modification of the Palatini action is another possible theory of
gravity in $4$ dimensions \cite{holst}. This action is in fact the starting 
point for Loop Quanum Gravity (LQG) (see \cite{cg_holst} for a detailed
comparison of the Palatini and the Holst action). It is also possible to
study the formalism of isolated horizons using the Holst action. Indeed,
it is possible to construct the covariant phase space of the Holst action
admitting a WIH as an inner boundary. Through this completely
covariant formulation, we also proved that the bounadry symplectic structure is
that of a $U(1)$ Chern-Simons theory \cite{cg_holst}. Entropy of WIH can be
calculated just as before by quantisation of the topological theory. Thus, if
the Holst action is further modified to include a non-minimal scalar
field coupled to gravity, the WIH
formulation will provide an ideal set-up to study the effects of these
non-canonical fields on black hole entropy. In other words, the Holst action
modified to include non-minimally scalar field and extended to WIH
formulation will precisely be able to tell
us, in a covariant framework, to what extent the scalar field contribute to the
entropy of black holes in these theories.

To proceed for such calculations, we need to go through a series of steps.
We shall first modify the boundary conditions for WIH making it amiable to the
case of non- minimal coupling. We will follow the modified boundary conditions
already stated in \cite{acs}. The zeroth law for the black holes
will simply follow from these conditions. Secondly, we shall have to modify
the Holst action to include the non- minimal scalar field coupling
\footnote{We shall use the acronym `non-minimally coupled Holst(Palatini)
action' to mean the Holst(Palatini) action modified to include that effect of 
scalar fields non-minimally coupled to gravity}. This is
non-trivial because we have to ensure that the phase space of this theory
can obtained from non- minimal Palatini theory \cite{acs} by a one parameter
canonical transformation (just like in the standard case of minimal
coupling). Then the covariant phase space for the non- minimally coupled Holst
action admitting a WIH as the internal boundary
needs to be constructed. The first law for the WIHs in this theory can be
proved using the symplectic structure constructed on this covariant phase space.
The first law for these horizons are expected to be modified
induced by the non- minimal scalar coupling.
This result was already obtained previously for the \emph{Weakly Isolated
Horizons} using the non- minimally coupled Palatini action.
We shall rederive these results using the non-minimal Holst action and for a
more general class of horizons.
This derivation of the first law (from the symplectic structure of
Holst action) will be non-trivial because the extra contrbutions form the
Holst action precisely cancel so as to lead to
the standard results from the Palatini theory \cite{acs}. We shall then show
that on the phase space containing spherical
horizons of fixed area and show that the surface symplectic structure acquires
the structure of a Chern-Simons theory although the scalar field will
appear as a lebel of the Chern-Simons theory (and hence the entropy).
This possibllity was already predicted from Killing Horizon framework
\cite{Wald}. Such a dependence on scalar field was also derived through
canonical formalism in \cite{ac}. We put this result on a firmer basis by
rederiving in a completely covariant way.

The plan of the paper is as follows. First section will contain a quick
introduction of the WIH formalism.
The zeroth law will also be proved in this section. The second section will be
used to define
the non- minimally coupled Holst action and then will be followed by
construction of the space of solution of the theory admitting a WIH as an
internal boundary.
The first law for the black holes in this theory will be caried out in the 
next section. We will also provide an alternative 
derivation of the first law. The fourth section will be devoted to the
derivation of the
Chern- Simons boundary symplectic structure on the phase space of fixed area.

\section{Weak Isolated Horizons}

We provide a brief introduction to Weak Isolated Horizons (WIH). (see \cite{cg,
cg_holst} for details). Let $\cal M$  be a four-manifold 
equipped with a metric $g_{ab}$ of signature $(-,+,+,+)$ and $\nabla_{a}$ be
the covariant derivative compatible with $g_{ab}$.
Consider a  null hypersurface ${\Delta}$ in $\cal M$. The surface $\Delta$
naturally admits an equivalence class of null normals 
$[\,\xi\ell^a\,]$, $\xi$ being any arbitrary positive function.
We denote by $q_{ab}\triangleq g_{\ub{ab}}$ the degenerate intrinsic metric on
$\Delta$ induced by $g_{ab}$\footnote{indices that are not explicitly intrinsic
on
$\Delta$ will be pulled back and $\triangleq$ means that the equality holds
{\em only on} $\Delta$}. The expansion
$\theta_{(\ell\,)}$ of the null normal $\ell^a$ is then defined by
$\theta_{(\ell\,)}=q^{ab}\nabla_a\ell_b$, where $\nabla_a$ is the covariant
derivative compatible with $g_{ab}$. We shall work with the null tetrad basis
$(\ell, n, m,
\bar{m})$ such that $1\!=\!-n\cdot\ell=\!m\cdot\bar m$ and all other scalar
products vanish. This is specially suited for the present problem since one of
the 
null normals $\ell^a$ matches with one of the vectors in the equivalence
class $[\xi\ell^a]$. The spacetime metric, in terms of this null basis is then
given by $g_{ab}=-2\ell_{(a}
n_{b)}+2 m_{(a} \bar m_{b)}$.

We shall now impose a  minimal set of boundary conditions on
the null surface $\Delta$ so that effectively the surface behaves as a black
hole horizon \cite{cg,cg_holst, acs}.  Since the null surface is generated
by 
an equivalence class of null normals $[\xi\ell^a]$, it is natural
to ensure that the boundary conditions hold for the entire equivalence class.
This would seem to imply that we need infinite number of boundary
conditions, one for each $\ell^a$ in $[\xi\ell^a]$. However, we shall see
that in a restricted class of the function $\xi$ (which we will specify in
this subsection, see eqn \ref{restclass}), the boundary conditions are such that
they are satisfied for any $\ell^a$ in $[\xi\ell^a]$ if they hold for
one $\ell^a$.

The null surface $\Delta$ generated by the equivalence class $[\xi\ell^a]$ will
be called a \textit{weak isolated
horizon} (WIH) in $({\cal M},g_{ab})$ if the following conditions are satisfied:
\begin{enumerate}
\item $\Delta$ is topologically $S^2 \otimes \mathbb{R}$.
\item The expansion $\theta_{(\xi\ell)}\=0$ for any $~\xi\ell^a$ in the
  equivalence class.
\item The equations of motion
 hold on the surface $\Delta$ and the non- minimally coupled scalar field
$\phi$ is such that $\lie_{\xi\ell}\phi\=0$.
\item There exists a one- form $\omega^{(\xi\ell)}$ on $\Delta$ so that
$\lie_{\xi\ell}\omega^{(\xi\ell)}\=0$ \label{wih}.
\end{enumerate}
All the boundary conditions, except the fourth one, hold for the entire
equivalence class
if they hold for one representative of the equivalence class $[\xi\ell^a]$.
Each null normal $\xi\ell^a$ is geodetic by construction 
\begin{equation}\label{eqgeodetic}
\xi\ell^a\grad_a(\xi\ell^b)\triangleq\kappa_{(\xi\ell\,)}\xi\ell^b
\end{equation}
where $\kappa_{(\xi\ell\,)}$ is the acceleration of $\xi\ell^a$. It is easy to
see from (\ref{eqgeodetic}) that the acceleration varies in the equivalence
class
\begin{equation}\label{kappa_eq_class}
\kappa_{(\xi\ell\,)}=\xi\kappa_{(\ell\,)}+\ellie_\ell\xi.
\end{equation}
The boundary conditions then imply that 
each null normal in the equivalence class $[\xi\ell^a]$ is  twist-free,
shear-free
and
\begin{equation}\label{omega_def}
\nabla_{\ub{a}}\ell^b\=\omega^{(\ell\,)}_{a}\ell^b
\end{equation}
where $\omega^{(\ell\,)}_{a}$ is a one-form on $\Delta$ associated with the null
normal $\ell^a$
which varies in the equivalence class as
\begin{equation}\label{omega_exp_eqclass}
{\omega}^{(\xi\ell\,)}= \omega^{(\ell\,)}+d\ln\xi,
\end{equation}
where $d$ is the exterior derivative in $\Delta$. It follows that each
$\xi\ell^a$ 
in the class is a \emph{Killing vector field} on $\Delta$, namely
$\ellie_{(\xi\ell\,)}\,q_{ab}\triangleq
0$. It also follows that the 
curvature of $\omega^{(\xi\ell\,)}$ is (see appendix of \cite{cg_holst} for
details)
\begin{equation} \label{curvomega}
d\omega^{(\xi\ell\,)}\triangleq 2({\rm Im}\Psi_2)\,{}^2\mbf{\epsilon},
\end{equation}
where ${\rm Im}\Psi_2=C_{abcd}\ell^am^b\bar m^cn^d$ is a complex scalar
associated with the Weyl-tensor
$C_{abcd}$ and ${}^2\mbf{\epsilon}=im\wedge\bar m$ is the 
area two-form of the cross-sections (these are $v=$ constant sections where $v$
is the affine parameter of $\ell^a$ such that $\ellie_{\ell}v=1$) of $\Delta$.
From
(\ref{omega_exp_eqclass}) it is obvious that (\ref{curvomega}) will hold for all
$\omega^{(\xi\ell\,)}$
in the class. Since each $\xi\ell^a$ is Killing, the area two-form
${}^2\mbf\epsilon$ is preserved under its Lie-flow
$\ellie_{(\xi\ell\,)}{}^2\mbf{\epsilon}_{ab}\triangleq 0$.

The fourth boundary condition is \emph{not} valid for any all null-normals.
This condition
can be viewed as a restriction
on the function $\xi$ so that in this restricted class, any null-normal in 
$[\xi\ell^a]$ will satisfy all the boundary conditions if it holds one.
We restrict the choice of $\xi$s to
\beq\label{restclass}
\xi=c\,e^{-v\kappa_{(\ell\,)}}+\kappa_{(\xi\ell\,)}/\kappa_{(\ell\,)},
\eeq
where $c$ is a nonzero function satisfying $\ellie_\ell c=0$
and $v$ is the affine parameter of $\ell^a$. In the rest of the paper we choose
$c\triangleq$ constant.
It is easy to see that the fourth boundary condition gives
\beq \ellie_{(\xi\ell\,)}\,\omega^{(\xi\ell\,)}\triangleq
d\kappa_{(\xi\ell\,)}\= 0\label{0thlaw}\eeq 
for any $\xi$ belonging to the restricted class (\ref{restclass}).
This is equivalent to the {\em zeroth law} which states that the
surface gravity associated with each $\xi\ell^a$ in the equivalence class
is constant on $\Delta$. The restricted class (\ref{restclass}) admits a
$\xi=c\,e^{-v\kappa_{(\ell\,)}}$ such that $\kappa_{(\xi\ell\,)}\triangleq 0$
when $\kappa_{(\ell\,)}\neq 0$.
For obvious reasons such a WIH will be called {\em extremal}. Thus, the
restricted class of
null normals, as opposed to the constant class of null normals $[\,c\ell^a\,]$,
contains both
extremal and non-extremal horizons. In other words, WIH boundary conditions are
sufficiently
weak to accommodate both types of horizons.

We shall work with spherical horizons so that
the scalar field is spherically symmetric. The 
boundary condition then imply that 
the field is constant on the horizon. The case for non- spherical horizons
will be dealt with elsewhere.

\section{Non- minimal Coupling of Holst Action and WIH}

The Holst action non-minimally coupled to a scalar field is given by the
following action:
\begin{equation}\label{nmin_holst1}
S[e,A, \phi] =
 \int_{\mathcal M} d^4\!x\,\,  e\, \left[
\frac{1}{16\pi G}\, f(\phi)\,  e^a_I  e^b_J \left(F( A)_{ab}^{IJ}
-\frac{1}{2\gamma}\epsilon_{KL}{}^{IJ}F( A)_{ab}^{KL}\right) 
- \frac{1}{2}\,K(\phi) \del_a \phi \del_b \phi\,\,
 e^a_I  e^b_J \eta^{IJ} - V(\phi)\right ] \,  ,
 \end{equation}
where
\be \label{G} K(\phi) = [1 + (3/16\pi G) (f'(\phi))^2/f(\phi)]\, .\ee
Here $e^a_I$ is the tetrad, $e$ its determinant, $F_{ab}
^{IJ}$ is the curvature of the connection $A_{a}^{IJ}$ , and $\gamma$
is a fixed but arbitrary number and is called the Barbero- Immirzi parameter.

Consider the case when the manifold has no inner boundary.
Variation with respect to $A$ yields the equation of motion for
the connection:
 \be
\nabla_{a}\,\left(f(\phi)\,  e\,   e^a_{[I}  e^b_{J]}\right)=0,
 \label{eq_compatible1}
\ee
where $\nabla_{a}$ is the covariant derivative operator 
corresponding to the connection $A_{IJ}$ and
acts both on the spacetime and the Lorentz indices.
The boundary contributions at the spatial infinity can be taken care from
the asymptotic flatness conditions. 
The equation (\ref{eq_compatible1}) has a dependence of the
scalar field. To solve this equation, assume that
the rescalings of the tetrads \cite{acs}:
\be
\hat e^a_I= ( p\,e^a_I)
\ee
where $p= 1/\sqrt{f(\phi)}$, is well-defined and
non-degenerate. The determinant $e$ is then also rescaled so that 
$p^{4}~\hat e = e$. It then follows that the equation (\ref{eq_compatible1})
can be rewritten as:
 \be\label{eq_compatible2}
 \nabla_a \,\left(\hat e\,   \hat e^a_{[I}  \hat e^b_{J]}\right)=0.
\ee
The form of the equation (\ref{eq_compatible2}) simply suggests that
 $A_{IJ}$ is the unique Lorentz spin
connection compatible with $\hat{e}^a_I$. 
Also, when the above equation of motion is satisfied, \emph{i.e} 
when the connection $A_{IJ}$ is the spin connection, the
second term in the action is precisely the Bianchi identity
and hence is zero. Furthermore, when the connection is a spin connection, the
the standard non- minimally coupled Einstein- Hilbert action can recovered 
upto a surface term \cite{acs}.
The extra non- minimally coupled Holst modification in fact induces
canonical transformation on the phase space labelled by $\gamma$.
While the symplectic structure can be shown to be immune to this canonical
transformation,
the quantum theory is sensitive to these $\gamma$ sectors.

\begin{figure}[h] \label{f1}
  \begin{center}
  \includegraphics[height=4.0cm]{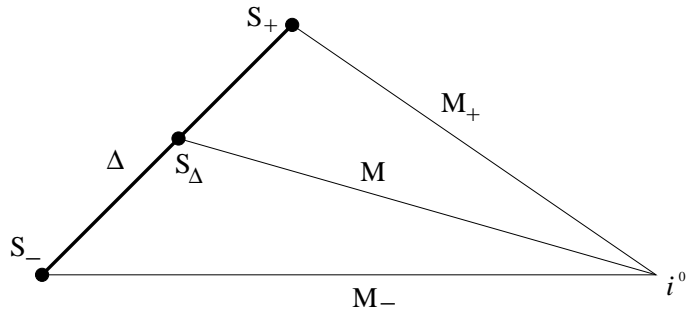}
  \caption{$M_\pm$ are two partial Cauchy surfaces enclosing a  region of
space-time and intersecting $\Delta$ in the $2$-spheres $S_\pm$ respectively
and extend to spatial infinity $i^o$. Another Cauchy slice M is drawn which
intersects $\Delta$ in $S_\Delta$}
  \end{center}
\end{figure}

\subsubsection*{Tetards and Connection on $\Delta$}\label{subsec:connection} 
In this subsection, we make a detailed study of the Holst action in a spacetime
region which is bounded by an inner boundary (a WIH) and two Cauchy 
surfaces $M_{+}$ and $M_{-}$ extended to spatial infinity (see fig. 1). The
variation of the fields is subject to the WIH boundary conditions on $\Delta$
and the asymptotic flatness conditions at spatial infinity.
For convenience let us choose a fixed set of internal null vectors $(\ell^{I},
n^{I}, m^{I}, \bar m^{I})$
on $\Delta$ such that $\partial_{a}(\ell^{I}, n^{I}, m^{I}, \bar m^{I})=0$ (this
partially fixes
the internal Lorentz frame). Given these internal 
null vectors and the tetrads $e^I{}_{a}$, we can construct the null vectors
$(\ell^{a}, n^{a}, m^{a}, \bar m^{a})$ through $\ell_a =e^I{}_{a}\ell_{I} $
etc.. 

In case WIH is an inner boundary, we must take appropriate care 
to verify the variational principle while the equation of motion are
determined. This requires expressions of the tetrads and the connection
pulled back to the WIH. 
On WIH $\Delta$, the expression of wedge product of tetrads is given by
\begin{equation}\label{tetrad_prod_exp}
e^{I}{}_{\ub{a}}\wedge e^{J}{}_{\ub{b}} \= -2~n_{a}\wedge m_{b}~\ell^{[I}\bar
m^{J]} -2~n_{a}\wedge \bar m_{b}~\ell^{[I} m^{J]} +2i~m^{[I}\bar
m^{J]}~{}^2\mbf{\epsilon}_{ab}
\end{equation}
Using this expression for the tetrad products (\ref{tetrad_prod_exp}), and the
expansion
for the internal epsilon tensor $\epsilon_{IJKL}=4!\ell_{[I}n_{J}m_{K}\bar
m_{L]}$, the 
expression for $\Sigma_{\ub{ab}}{}^{IJ}$, restricted to $\Delta$ is
\begin{equation}\label{sigma_tetrad_exp}
\Sigma_{\ub{ab}}{}^{IJ}\= 2\ell^{[I}n^{J]}~ {}^2\mbf{\epsilon}_{ab}
+2n_{a}\wedge(im_{b} \ell^{[I}\bar m^{J]} - i\bar m_{b}\ell^{[I} m^{J]})
\end{equation}

To find an expression for the connection $A_{IJ}$ on $\Delta$, we need the
covariant derivatives
of null normals pulled back and restricted to $\Delta$. We shall use the Newman-
Penrose formalism (see appendix of \cite{cg_holst} for details ).
Note that the covariant derivative does not
annihilate the metric $g_{ab}$, but the conformal metric $p^{-2}g_{ab}$.
However, the inner product is still taken with respect to the
metric $g_{ab}$, see eqn. \ref{eq_compatible2}.
This leads to a modification of the connection compared to the standard 
minimally coupled case, as is collected below:

\begin{equation} \nabla_{\ub{a}}\ell^{b}\= \omega^{(\ell\,)}_{a}\ell^{b}
\end{equation}
\begin{equation}\label{exp_normal_n}
\nabla_{\ub{a}}n^{b}\=-(\omega^{(\ell\,)}_{a}+2~\nabla_{a}p)
n^{b}+\bar{U}^{(l,m)}_{a}m^{b}+U^{(l,m)}_{a} \bm^{b}
\end{equation}
\begin{equation} \nabla_{\ub{a}}m^{b}\={U^{(l,m)}}_{a}\ell^{b}+ V^{(m\,)}_{a}
m^{b}
\end{equation}
\begin{equation}
\nabla_{\ub{a}}\bm^{b}\=\bu^{(l,m)}_{a}\ell^{b}-(V^{(m\,)}_{a}+2~\nabla_{a}
p)\bm^{
b}
\end{equation}
where, the superscripts for each of the one forms keep track of their
dependencies on the rescaling of the corresponding null normals. The expressions
of the one forms $\omega^{(\ell\,)}, U^{(l,m)}, \bu^{(l,m)}$ and $V^{(m)}$ can
be written in terms of the null normals and are as follows:
\begin{eqnarray}\label{definitions_NP} \omega^{(\ell\,)}_{a}&\=& -\left(
\epsilon + \bepsilon\right)n_{a}+\left(
\balpha+\beta \right)\bm_{a}+\left( \alpha+\bbeta \right)m_{a}\nn 
U^{(l,m)}_{a} &\=&-\bpi n_{a}+\bmu m_{a} +\blambda \bm_a \nn
 V^{(m\,)}_{a}&\=&-\left( \epsilon-\bepsilon\right)n_{a}+\left( \beta -\balpha
\right)\bm_{a}+\left(\alpha-\bbeta \right)m_{a}
\end{eqnarray}
The part of the connection $V^{(m\,)}$ is purely imaginary.

We can use these information to find the connection. 
Since the internal null vectors are fixed,  we get 
\begin{equation}\label{eq-conn-cal}
\nabla_{\ub{a}}\ell_{I}\= A^{(\ell)}_{\ub{a}I}{}^{J}\ell_{J}.
\end{equation}
This tetrad is annihilated by the covariant derivative,
$\nabla_{a}e_{b}^{I}=0$. 
Then the equation (\ref{eq-conn-cal}) gives:
$A_{\ub{a}}{}^{I}{}_{J}\ell^{J}\=\omega_{a}^{(\ell)}\ell^{I}$.
Written is a more compact form, this reduces to
\begin{equation}
 A^{(\ell)}_{\ub{a}IJ}\=-2(\omega_{a}^{(\ell)}+\nabla_{a}p)~\ell_{[I}n_{J]} +
Q_{IJ},
\end{equation}
where, the one form $ Q_{IJ}$ is such that $ Q_{IJ}\ell^{J}\=0$.
This construction can be followed for other null vectors $n_{I},~
m_{I}$ and $\bar m_{I}$. 
The connections obtained for these other null vectors $n$, $m$,
$\bm$ will complement each other.
Combining all the expression for these connections, we get the complete
expression for the connection $A_{\ub{a}IJ}$
\begin{equation} \ub{A}_{IJ}\=-2~(\omega^{(\ell)}+\nabla_{a}p)~ \ell_{[I}n_{J]}+
2~U^{(l,m)}~\ell_{[I}\bm_{J]}+ 2~{\bar U}^{(l,m)}~\ell_{[I}
m_{J]}+ 2~(V^{(m)}+\nabla_{a}p)~m_{[I}\bm_{J]}
\end{equation}
We define the following connection for ease of computation 
\begin{equation} 
A^{(H)}_{IJ}:=\frac{1}{2}\left( A_{IJ}-\frac{\gamma}{2}
\epsilon_{IJ}{}^{KL}A_{KL}\right)
\end{equation}
This leads to the following form of the connection:
\begin{eqnarray} A^{(H)}_{\ub{a}IJ}&\=& ~\ell_{[I}n_{J]}~\left(
-\omega^{(\ell)}_{a}+i\gamma V^{(m)}_{a}
\right) + ~ m_{[I}\bm_{J]}~\left(V^{(m)}_{a}-i\gamma \omega^{(\ell)}_{a}
\right)\nn
&+& ~\ell_{[I}\bm_{J]}~\left( U^{(l,m)}_{a}+i\gamma U^{(l,m)}_{a}
\right)+ ~\ell_{[I}m_{J]}~\left(\bu^{(l,m)}_{a}-i\gamma \bu^{(l,m)}_{a}
\right)\nn
&-&(1-i\gamma)\nabla_{a}p\left(\ell_{[I}n_{J]}-m_{[I}\bm_{J]}\right )
\end{eqnarray}

Let us at this stage
point out the result of the rescaling of the null normal $\ell^a $ on the
various quantities of interest. Firstly, for $\ell^a\longrightarrow  \xi\ell^a$,
we have:
\begin{equation}
\omega^{(\ell\,)}_{\ub{a}}\rightarrow
\omega^{(\xi\ell\,)}_{\ub{a}}=\omega^{(\ell\,)}_{a}+\nabla_{\ub{a}}ln \xi
\end{equation} 
Since the normalization of $\ell^a$ and $n^a$ are connected, we must have
$n^a\longrightarrow  \frac{n^a}{\xi} $ when $\ell^a\longrightarrow 
\xi\ell^a $.
Then the effect of the rescaling can be seen to be: 
\begin{equation}
\nabla_{\ub{a}}\left( \frac{n^{b}}{\xi}\right)
\=-\omega^{(\xi\ell\,)}_{a}\left(
\frac{n^{b}}{\xi}\right) + \bar{U}^{(\xi\ell,m)}_{a}
m^{b}+ U^{(\xi\ell,m)}_{a}\bm^{b}
\end{equation}
Thus, under this transformation, the components of the connection transform as 
\begin{eqnarray} 
&&\omega^{(\ell\,)}\rightarrow
\omega^{(\xi\ell\,)}_{a}\=\omega^{(\ell\,)}_{a}+\nabla_{\ub{a}}ln \xi 
,\nonumber \\
&&\bar{U}^{(\ell,m)}_{a}\longrightarrow
\bar{U}^{(\xi\ell,m)}_{a}\=\frac{\bar{U}^{(\ell,m)}_{a}}{\xi}~~~~
\mbox{and}\, ~~ U^{(\ell,m)}_{a}\longrightarrow
U^{(\xi\ell,m)}_{a}\=\frac{U^{(\ell,m)}_{a}}{\xi}
\end{eqnarray}

We can also independently rescale 
the other set of null vectors $m,\bar m$ of the
null tetrad.
This rescaling is completely
free of any information about the rescaling in the $\ell, n$  sector. Since
$\xi$
recsaling controls the extremality of the horizon, this means that whatever be
the nature of the horizon (extremal or non-
extremal) rescaling of $m, \bar m$ is always possible. Let, 
$m\rightarrow f m$ and  $\bm\rightarrow \frac{\bm}{f}$, 
where $f$ is any function on $\Delta$. Note that this implies that $f$ must
be a pure phase of the form $e^{i\Psi}$. The tranformations are
\begin{eqnarray}
\nabla_{\ub{a}}\left(f m^{b}\right)&\= &U^{(\ell,fm)}
_{a}\ell^{b}+V^{(fm\,)}_{a}\left(fm^{b}\right)\nonumber \\
\nabla_{\ub{a}}\left(\frac{\bm}{f}\right)&\=
&\bar{U}^{(\ell,fm)}_{a}\ell^{b}-V^{(fm\, )
}_{a}\left(\frac{{\bm}^{b}}{f}\right)
\end{eqnarray}
The transformation rules are as follows for the one forms $U^{(\ell,m)}_{a}$,
$\bar{U}^{(\ell,m)}_{a}$ and $V^{(m\,)}_{a}$ are
as follows:
\begin{eqnarray}\label{trans_rule_uuv} 
&&\bar{U}^{(\ell,m)}_{a} \longrightarrow
\bar{U}^{(\ell,fm)}_{a}\=\frac{\bar{U}^{(\ell,m)}_{a}}{f}\\ \nonumber
&&U^{(\ell,m)}_{a}\longrightarrow U^{(\ell,fm)}_{a}\=f U^{(\ell,fm)}_{a}\\
\nonumber
&&V^{(m\,)}_{a}\longrightarrow V^{(fm\,)}_{a}\=V^{(m\,)}_{a}+ \nabla_{\ub{a}}
~lnf
\end{eqnarray}

The part of the connection $\omega^{(\ell)}$ and $V^{(m\,)}$ transform
as abelian gauge field whereas the the other parts of connections 
only rescale.

\subsubsection*{Variation of the Action}
We take the Lagrangian $4$-form appropriate to the action of the Holst action to
be 
\begin{eqnarray}\label{lag4form}
-16\pi G\gamma~L &=& \gamma f(\phi)\Sigma_{IJ}\wedge F^{IJ}~-~f(\phi)e_{I}\wedge
e_{J}\wedge F^{IJ}~ -~\gamma~d(f(\phi)\Sigma_{IJ}\wedge A^{IJ})
+~d(f(\phi)e_{I}\wedge e_{J}\wedge A^{IJ})\nn
&-& 8\pi G ~K(\phi) {}^\star d\phi\wedge
d\phi +16\pi G ~V(\phi)\epsilon ,
\end{eqnarray}
where we have added the two boundary term just for convenience. They will
not contribute
to the equation of motion but will affect the boundary terms that will arise in
the subsequent
derivation.
The variation of the action on-shell will give two terms on the 
boundary $\Delta$. They are:
\begin{equation}
\delta S (e,A) = \frac{-1}{8\pi G \gamma}\int_{\Delta}\left[(iV^{(m)} +
\gamma~\omega^{(\ell)}) +
(1+\gamma)~dp\right]\wedge\delta(f(\phi){}^2\mbf{\epsilon})
\end{equation}
We will argue that the term is zero and hence the action principle is well
defined. The nature of argument is almost similar to that in \cite{cg_holst}.
We however repeat the arguments for the sake of completeness.
First of all, the field configurations
over which the variations are taken are such that they satisfy the standard
boundary conditions at infinity and the WIH boundary conditions at $\Delta$.
This immediately implies that the scalar field
does not affect the variation of the action. We only have to worry about the 
other terms. The weak isolation condition implies that
$\lie_{\ell}\omega^{(\ell)}\=0$
though there is no such condition on $V^{(m)}$. However 
interestingly, $d\omega^{(\ell)}$ and $dV^{(m)}$ are proportional
to ${}^2\mbf{\epsilon}$ and hence inner product with $\ell^a$ of these 
quantities are zero. This implies that for variations among
field configurations with null normals in the equivalence class,
we have $\lie_{\xi\ell}\omega^{(\ell)}\= d(\xi\kappa_{(\ell)})$ and
$\lie_{\xi\ell}V^{(m)}\= d(\xi(\epsilon-\bar\epsilon))$. 
This implies that on the application of  $\lie_{\xi\ell}$, the integral goes
to the initial and the final cross section of $\Delta$. However, the variation
of the fields for example $\delta{}^2\mbf{\epsilon}$
is zero at the initial and final hypersurface by the standard rules of
variational principle. Thus the integral is lie dragged by any null normal in
the equivalence class. In other words, the integral in zero at the initial and
the final hypersurface and is lie dragged on $\Delta$ . Thus, the entire
integral is zero and the action principle is well defined.

\subsubsection*{The Symplectic Structure}
\label{sec:Symplectic}
The construction of the symplectic structure from a given Lagrangian is
detailed in \cite{abrlw}. One first
extracts the symplectic one-form  $\Theta$ (spacetime three-form in
$4$-dimensions)
from the variation of the Lagrangian such that $\delta L=
d\Theta (\delta)$ where $\delta$ is an arbitrary vector field in the phase
space. In the present case, we have
\begin{equation}
16\pi G\gamma~\Theta(\delta)
% \gamma~\delta(f(\phi )\Sigma_{IJ})\wedge A^{IJ}-\delta(f(\phi )e_{I}\wedge
% e_{J})\wedge A^{IJ}
=-2~\delta(f(\phi)e^{I}\wedge e^{J})\wedge A^{(H)}_{IJ}+
K(\phi)~{}^{*}d\phi~\delta\phi
\end{equation}
The construction of the symplectic current from here is standard.
The current is $J(\delta_{1}, \delta_{2}):=\delta_{1}\Theta(\delta_{2})-
\delta_{2}\Theta(\delta_{1})$.
The current is closed on- shell \emph{i.e.} $dJ=0$.
 The resulting Symplectic Current is :
\begin{equation} J\left( \delta_1, \delta_2
\right):=\dfrac{1}{8\pi G\gamma}\left\lbrace \delta_{[1}\left(
f(\phi)~e^{I}\wedge
e^{J}\right) \right\rbrace \wedge\left\lbrace \delta_{2]}\left(
A_{IJ}-\frac{\gamma}{2} \epsilon_{IJ}{}^{KL}A_{KL}\right)\right\rbrace
-K(\phi)~\left\lbrace
\delta_{[1}\left({}^{*}d\phi\right)~\delta_{2]}\phi\right\rbrace
\end{equation}
Since $dJ=0$, upon integrating the symplectic current over $\mathcal{M}$, we get
contributions only
from the boundaries under consideration
\begin{equation}
\int_{M_{+}\cup M_{-}\cup\Delta\cup i^{0}}J(\delta_{1}, \delta_{2})=0
\end{equation}
The boundary conditions at infinity ensure that the integral of the symplectic
current
at spatial infinity vanishes. To construct the symplectic structure we
must be
careful that no data flows out of the phase space because of our choice of
foliation.
In other words, the symplectic structure should be independent of the
choice of foliation. To this end, we introduce potentials,
\begin{enumerate}
\item $\lie_{(\xi\ell)}\psi_{(\xi\ell)}\=\xi\ell^a
\omega^{(\xi\ell)}_{a}\=\kappa_{(\xi\ell)}$
\item $\lie_{(\xi\ell)}\mu_{(m)}\=i\xi\ell^a V^{(m)}_{a}\=
i\xi(\epsilon-\bar\epsilon)$
\end{enumerate}
which satisfy the boundary conditions that they are zero at the initial
cross-section
of $\Delta$ so that the additive ambiguities in them are removed. We choose
$\psi_{(\xi\ell)}=0$ and $\mu_{(m)}=0$ at $S_{-}$.
This potentials imply that $J(\delta_{1}, \delta_{2})\= dj(\delta_{1},
\delta_{2})$
\begin{equation}\label{exp_j_delta_d}
J(\delta_{1}, \delta_{2}))|_{{}_{\Delta}}\=d\left[\frac{-1}{8\pi G\gamma}
\left(\delta_{1}(f(\phi)~{}^2\mbf{\epsilon})~\delta_{2}(\mu_{(m)}
+\gamma\psi_{(\ell)})- (1\leftrightarrow 2)\right)\right]
\end{equation}
With this simplification, the integrals of $J(\delta_{1}, \delta_{2})$ on
$\Delta$
will be taken to the boundaries $S_{\pm}$ of $\Delta$.
We take a particular orientation of the spacetime foliation into account
and get
\begin{equation}
(\int_{M_{+}}- \int_{M_{-}})J(\delta_{1}, \delta_{2})\=\frac{1}{8\pi
G\gamma}(\int_{S_{-}}-\int_{S_{+}})\{\delta_{1}(f(\phi)~{}^2\mbf{\epsilon}
)~\delta_{2}(\mu_{(m)} +\gamma\psi_{(\ell)})- (1\leftrightarrow 2)\}
\end{equation}
The construction of symplectic current is independent of our choice foliation
and hence all the phase space information can be obtained from this symplectic
current
by staying on any arbitrary foliation. 
We choose a particular Cauchy surface $M$ which intersects $\Delta$
in the sphere $S_{\Delta}$ so that
\begin{eqnarray}\label{Palatini_1}
\Omega(\delta_{1}, \delta_{2}
)&:=&\frac{1}{8\pi G\gamma}\int_{M}\left[ \delta_{1}(f(\phi)~e^{I}\wedge
e^{J})~\wedge\delta_{2}A^{(H)}_{IJ} -\delta_{2}(f(\phi)~e^{I}\wedge
e^{J})~\wedge\delta_{1}A^{(H)}_{IJ} \right] \nn
&+&\frac{1}{8\pi G\gamma}\int_{S_{\Delta}}\left[
\delta_{1}(f(\phi)~{}^2\mbf{\epsilon})~\delta_{2}(\mu_{(m)}+
\gamma\psi_{(\ell)})
-
\delta_{2}(f(\phi)~{}^2\mbf{\epsilon})~\delta_{1}(\mu_{(m)} +
\gamma\psi_{(\ell)})\right]\nn
&+&\int_{M} K(\phi)~\left[\delta_{1}({}^{*}d\phi)~\delta_{2}
\phi-\delta_{2}({}^{*}d\phi)~\delta_{1}\phi\right]
\end{eqnarray}
The symplectic structure (\ref{Palatini_1}) obtained from the non-minimally
coupled Holst action has some additional terms compared to that obtained form
the non-minimally coupled Palatini theory \cite{acs}. The Palatini symplectic
structure can be read-off by collecting the $\gamma$ independent terms. We
shall see that some miraculous cancellations among the $\gamma$ dependent terms 
lead to the usual result of first law as is obtained from the Palatini theory.

\section{The First Law }
\label{sec:phase_space} 
We have already stressed that WIH is a local definition of horizon unlike the 
event horizon or Killing horizon. The first law for event horizons studies
variations of quantities defined (or normalised) with respect to spatial
infinity. In the present case, we want the first law 
to relate variations of local quantities that are defined only at the horizon
without any reference to the rest of the spacetime. In other words, we
expect that the first law based on the definition of WIH should involve
only locally defined quantities. For WIH, surface gravity $\kappa_{(\xi\ell)}$
has already been defined at the horizon. We now must define energy locally. In
spacetime, energy is associated with a timelike Killing vector field. Given any
timelike vector field $W^a$ in spacetime, it naturally induces a vector field
$\delta_{W}$ in the phase space. The phase space vector field $\delta_{W}$ is
the generator of time translation in the phase space. If time translation is a
canonical transformation in the phase space then $\delta_{W}$ defines
a Hamiltonian function $H_{W}$ for us. The vector field $\delta_{W}$ is
\emph{globally Hamiltonian} if and only
if $X_{W}(\delta)=\Omega(\delta,\delta_{W})=\delta H_{W}$ for any vector field
$\delta$ in the phase space. On WIH, the vector
fields $W^a$ are restricted by the condition that it should be tangential on
$\Delta$. Just like the usual advanced time coordinate, analog of `time' 
translation on WIH is along the null direction (two other are
spacelike). It is generated by the vector field
$[\xi\ell^a]$. For global solutions 
this null normal vector field becomes timelike outside the horizon and is
expected to match with the asymptotic
time-translation for asymptotically flat spacetimes. 

We want to find out if the flow generated by the phase space vector
field $\delta_{\xi\ell}$ is
Hamiltonian. The action of the phase space vector
field  $\delta_{\xi\ell}$
on tensor fields is the lie flow $\ellie_{\xi\ell}$ generated by the vector
field
$\xi\ell^a$. For the above symplectic structure, $X_{(\xi\ell)}(\delta)$ gets
contribution from both the bulk and the surface symplectic structure. The bulk
term, thanks to
the equation of motion satisfied by the fields and their variations, contributes
only through the boundaries of the Cauchy surface $M$, which are the $2-$
spheres $S_{\Delta}$
and $S_{\infty}$ respectively: 

%%%
\begin{equation}\label{bulk_X}
X_{\xi\ell}(\delta)|_{M}=\frac{-1}{8\pi G}\xi\kappa_{(\ell)}\delta (f(\phi)~
\mathcal{A}_{\Delta}) -\frac{i}{8\pi G\gamma}\int_{S_{\Delta}}\xi(\epsilon
-\bar\epsilon)\delta (f(\phi)~ {}^2\mbf\epsilon) +  \delta E_{(\xi\ell)}
\end{equation}
where, $\mathcal{A}_{\Delta}=\int_{S_{\Delta}}\epsilon$ is the area of 
$S_{\Delta}$ and $E_{(\xi\ell)}$ is the ADM energy arising 
out of the integral at $S_{\infty}$, assuming that the asymptotic time 
translation matches with the vector field $\xi\ell^a$ at infinity.

The one- form  $X_{(\xi\ell)}(\delta)$ also gets contribution from the 
surface symplectic structure (the arguments are similar to that in \cite{cg,
cg_holst}). The action of $\delta_{(\xi\ell)}$ cannot
be interpreted as $\ellie_{(\xi\ell)}$ when acting on potentials.
To determine the action, we proceed as follows.
For the case of $\psi_{(\ell)}$, it is clear that since
variation of $\psi_{(\xi\ell)}$ is completely determined by
$\kappa_{(\xi\ell)}$, $\delta_{(\xi\ell)}\psi_{(\xi\ell)}=0$.
However, $\psi_{(\xi\ell)}=\psi_{(\ell)}+ln~\xi $ implies that
$\delta_{(\xi\ell)}\psi_{(\ell)}=-\ellie_{\ell}\xi$.
For the other potential, observe that
$\delta_{(\xi\ell)}\mu_{(m)}-i(\epsilon-\bar\epsilon)$
satisfies the differential equation
$\ellie_{\xi\ell}(\delta_{(\xi\ell)}\mu_{(m)}-i(\epsilon-\bar\epsilon))=0$
with the boundary condition that $\mu_{(m)}=0$ at the point $v=0$. This implies
that
because $(\epsilon-\bar\epsilon)=0$ at $v=0$, the action is
$\delta_{(\xi\ell)}\mu_{(m)}=i(\epsilon-\bar\epsilon)$. The considerations above
leads to
\begin{equation}\label{bndy_X}
X_{\xi\ell}(\delta)|_{S_{\Delta}}=
=-\frac{1}{8\pi G}\ellie_{\ell}\xi~\delta (f(\phi)~\mathcal{A}_{\Delta})
+\frac{i}{8\pi G\gamma}\int_{S_{\Delta}}\xi(\epsilon -\bar\epsilon)\delta
(f(\phi)~{}^2\mbf\epsilon)
%&=&-\frac{1}{8\pi G}\kappa_{(\xi\ell)}\delta \mathcal{A}_{\Delta}+\delta
\end{equation}

Combining the two equations (\ref{bulk_X}) and (\ref{bndy_X}), we get:
\begin{equation}
X_{\xi\ell}(\delta)\=-\frac{1}{8\pi G}\kappa_{(\xi\ell)}\delta
(f(\phi)~\mathcal{A}_{\Delta})+\delta E_{(\xi\ell)}
\end{equation}
For $\delta_{\xi\ell}$ to be Hamiltonian,
the surface gravity $\kappa_{(\xi\ell)}$ must be a function
of area $\mathcal{A}_{\Delta}$ only. This is reasonable since the phase
space is characterized by area (and charges) and so $\kappa_{(\xi\ell)}$ can
only
be a function of these quantities. The exact functional dependence of $\kappa$
on area is undetermined.
This is a fundamental result of the generalization to the generalised class of
null normals $[\xi\ell^a]$. In the constant class of null normals, there
is no contribution from the surface symplectic structure. In the
present case, the precise contribution (and cancellations)
from the bulk (\ref{bulk_X}) and the boundary (\ref{bndy_X}) leads to
the physically meaningful variation.

Previous results also imply that there exists
a locally defined function $ E_{\Delta}$ such that.
\begin{equation}\label{1stlaw}
\delta E_{\Delta}\= \frac{1}{8\pi G}\kappa_{(\xi\ell)}~\delta
(f(\phi)~\mathcal{A}_{\Delta})
\end{equation} 
such that $H_{\xi\ell}=E_{(\xi\ell)}-E_{\Delta}$ where $H_{\xi\ell}$ is the
associated
Hamiltonian function $X_{\xi\ell}(\delta)=\delta H_{\xi\ell}$. We shall
interprete $E_\Delta$ as the locally defined energy of the WIH and
(\ref{1stlaw}) as the first law of the WIH. $H_{\xi\ell}$ 
receives contributions both from the bulk as well as the boundary symplectic
structures has information of energy of the region between the WIH
and spatial infinity. The ADM energy $E_{(\xi\ell)}$ is the sum total of these
two energies. The explicit form of $E_\Delta$ can be determined iff 
the functional dependence of $\kappa_{(\xi\ell)}$ on area is
known.

\subsection*{Inclusion of rotation}
\label{rotfirstlaw}
A WIH is spherically symmetric if the geometry of the $v=$ constant ($v$ is the
affine parameter of $\ell^a$) sections of $\Delta$ is spherical. So in addition
to
$\xi\ell^a$, such horizons admit three other local spacelike Killing vector
fields
that are tangential to the cross-sections. To include rotations, we 
consider the horizon to have a symmetry
about some axis $\varrhoup^a$. For spherical symmetry,  $\varrhoup^a$
is one of the three Killing vectors. 
The metric for this case will be given by
\begin{equation}
 ds^2 = r_{\Delta}^{2}\left( d\varthetaup^{2} +
F(\varthetaup)^{2}d\varrhoup^{2}\right)
\end{equation}
%
% In this case, the relevant forms are
% %
% \begin{eqnarray}\ellabel{rotforms}
% m_{a}=\frac{r_{\Delta}}{\sqrt{2}}\elleft[ \elleft( d\varthetaup)\right) _{a} +
% iF(\varthetaup) \elleft( d\varrhoup\right)_{a}\right]  \nn
% \bar{m}_{a}=\frac{r_{\Delta}}{\sqrt{2}}\elleft[ \elleft( d\varthetaup)\right)
%_{a}
-
% iF(\varthetaup) \elleft( d\varrhoup\right)_{a}\right]
% \end{eqnarray}
%
where, $F(\varthetaup)=sin \varthetaup$ for spherical symmetry. We shall
keep the function as $F(\varthetaup)$ for further calculations. The
corresponding vector fields can also be found out easily. They are given
by:
\begin{eqnarray}\label{rotvecfields}
m^{a}=\frac{1}{r_{\Delta}\sqrt{2}}\left[
\left(\frac{\partial}{\partial\varthetaup}\right) ^{a} +
\frac{i}{F(\varthetaup)}\left(\frac{\partial}{\partial\varrhoup}\right)^{a}
\right] \nn
\bar{m}^{a}=\frac{1}{r_{\Delta}\sqrt{2}}\left[
\left(\frac{\partial}{\partial\varthetaup}\right) ^{a} -
\frac{i}{F(\varthetaup)}\left(\frac{\partial}{\partial\varrhoup}\right)^{a}
\right]
\end{eqnarray}

The volume element of the sphere is
$\epsilon=im\wedge{m}=r_{\Delta}^{2}F(\varthetaup)\left(d\varthetaup\wedge
d\varrhoup\right)$.
Let us now turn to the symplectic structure. The symplectic structure of the
non- minimally coupled Holst action (see \ref{Palatini_1}) admits a canonical
transformation to the symplectic
structure of the non- minimally coupled Palatini action. We shall not show this
explicitly, this can be derived fairly easily from the results in appendix
of \cite{cg_holst} 
\begin{eqnarray}\label{symplectic} 
\Omega(\delta_1,\delta_t)&=&\frac{1}{16\pi
G}\int_M\big[\delta_2(f(\phi)~\Sigma^{I\!J})\wedge
\delta_1A_{I\!J}-\delta_1(f(\phi)~\Sigma^{I\!J})\wedge\!\delta_2A_{I\!J}\big]\nn
&-&\frac{1}{8\pi G}\oint_{S_\Delta}\big[
\delta_2(f(\phi)~\mbf{\epsilon})\,\delta_1\psi_{(\ell\,)}-\delta_1
(f(\phi)~\mbf{\epsilon})\,\delta_2\psi_{(\ell\,)}\big]\;.
\end{eqnarray}
As in the derivation of the first law, the Killing vector $\varrhoup^{a}$
on the spacetime induces a vector field $\delta_{\varrhoup}$ on the phase space.
We now ask wheather the flow generated by $\delta_{\varrhoup}$ 
on the phase space is a
Hamiltonian. In that case, we can call the Hamiltonian function as angular
momentum.
In the symplectic strucure, the only contribution comes from the bulk
symplectic structure. The surface symplectic structure has no
contribution. This can be shown as follows:
First, the area element is Lie-dragged by the vector field $\varrhoup$. The
other
term in the symplectic structure is also zero by the following argument. We
know that
$\delta_{\varrhoup}\kappa_{(\ell)}=0=\delta_{\varrhoup}(\ell.\nabla\psi)$. This
implies that
$[\varrhoup,\ell].\nabla\psi+\ellie_{\ell}(\delta_{\varrhoup}\psi)=0$.
But the
first term is zero by the conditions of symmetry and to keep the foliation
fixed. This then implies that
$\ellie_{\ell}(\delta_{\varrhoup}\psi)=0$, \emph{i.e.} the function
$(\delta_{\varrhoup}\psi)$ is
a constant along $\ell$. However, since $(\delta_{\varrhoup}\psi)=0$ on $S_{-}$,
this implies that $(\delta_{\varrhoup}\psi)=0$ on each $S_{\Delta}$ on $\Delta$.
This result holds true for all $\ell^a$ in the class $[\xi\ell^a]$.

Proceeding as in previous sections, the bulk
symplectic structure contributes only through the boudaries of $M$ which are at
$\Delta$ and at infinity. The contribution of the bulk symplectic structure
to the bulk of $M$ vanishes due to the equation of motion satisfied by the
fields $(e, A)$ and the linearised equations satisfied by them. So, we only
get contribution from the surface term
of the bulk symplectic structure. Thus, the
symplectic structure reduces to:
\begin{equation}\label{rotsymplectic}
\Omega\left(\delta,\delta_{\varrhoup} \right)= \frac{-1}{8\pi
G}\oint_{S_{\Delta}}\left\lbrace \left[ \varrhoup\rfloor\omega\right]
\delta(f(\phi)~\mbf{\epsilon}) -f(\phi)\left[\varrhoup\rfloor\epsilon\right]
\wedge\delta\omega\right\rbrace
\end{equation}
We can determine the terms in the symplectic struture explicitliy. The first
term is found out as follows:
\begin{eqnarray}\label{values}
 \frac{\partial}{\partial\varrhoup}\rfloor\omega&=&\frac{ir_{\Delta}
F(\varthetaup)}{ \sqrt{2}}
\left( \pi-\bar\pi\right); \nn
\left[\frac{\partial}{\partial\varrhoup}\rfloor\omega\right] \delta\epsilon&=&
\frac{i}{\sqrt{2}}\left( \pi-\bar \pi)(2\delta r_{\Delta}F +r_{\Delta}\delta
F)\right)
\epsilon\nn
\end{eqnarray}

The second term in the expression of symplectic structure is also determined in
a similar fashion. The result is:
\begin{eqnarray}\label{values}
~~~~ \frac{\partial}{\partial\varrhoup}\rfloor
\epsilon&=&-r_{\Delta}^{2} F(\varthetaup) d\varthetaup\nn
\left[ \frac{\partial}{\partial\varrhoup}\rfloor\epsilon\right]
\wedge\delta\omega &=& \frac{-i}{\sqrt{2}}\delta\left[ r_{\Delta}F (\pi -\bar
\pi)\right] \epsilon ,
\end{eqnarray}
where, we have used the decomposition of $\omega$ in terms of the null normals
(see eqn \ref{definitions_NP}). Putting these relations in the symplectic
structure (eqn \ref{rotsymplectic}) and taking into
account of the fact that  $r_{\Delta}F\delta\epsilon=\left( 2\delta
r_{\Delta}F+
r_{\Delta}\delta F\right)\epsilon$, we get
\begin{equation}\label{redrotsymplectic}
 \Omega\left(\delta,\delta_{\varrhoup} \right)= \frac{-1}{8\pi
G}\oint_{S_{\Delta}}\delta\left\lbrace \frac{if(\phi)}{\sqrt{2}}(\pi - \bar
\pi)r_{\Delta}F \epsilon\right\rbrace
\end{equation}
Note that the symplectic structure is now a total variation. This is 
precisely the neccesary and sufficient condition for which there exists a 
 Hamiltonian vector field $\delta_{\varrhoup}$, \emph{i.e.} the vector field is 
a phase space
symmetry ($\ellie_{\delta_{\varrhoup}}\Omega=0$, everywhere on $\Gamma$). Then
one
can define
a function $(J^{(\varrhoup)}_{\Delta})$ which will generate diffeomorphism along
the
particular vector field $\varrhoup^a$  such that for all vector fields $\delta$
on $\Gamma$,
\begin{equation}
\delta J^{(\varrhoup)}_{\Delta}=\Omega(\delta,\delta_{\varrhoup})
\end{equation}
Using the expressions in the  eqn.(\ref{values}) and the expression for
$\epsilon$ and
$\omega$, we get
\begin{eqnarray}\label{omega_phi}
\Omega_{B}\left(\delta,\delta_{\varrhoup} \right)= \frac{-1}{8\pi
G}\oint_{S_{\Delta}}\delta\left[f(\phi) \left(
\frac{\partial}{\partial\varrhoup}\rfloor\omega\right) \epsilon\right] \nn
=\frac{1}{8\pi G}\oint_{S_{\Delta}}\delta\left[ \left(f(\phi)
\frac{\partial}{\partial\varrhoup}\rfloor\epsilon\right)\wedge\omega \right] ,
\end{eqnarray}

where the second term in the above expression eqn. (\ref{omega_phi}) is obtained
by
noting that
$(\varrhoup\rfloor\omega)\epsilon=-(\varrhoup\rfloor\epsilon)\wedge\omega$.
The angular momentum at the horizon corresponding to the vector field
$\varrhoup^{a}$ is thus defined to be
\begin{eqnarray}
 J^{(\varrhoup)}_{\Delta}&=&\frac{1}{8\pi
G}\oint_{S_{\Delta}}\left(f(\phi)~\frac{\partial}{\partial\varrhoup}
\rfloor\epsilon
\right)\wedge\omega \nn&=&\frac{-1}{8\pi G}\oint_{S_{\Delta}}\left[
f(\phi)~\left(
\frac{\partial}{\partial\varrhoup}\rfloor\omega\right) \epsilon\right]
\end{eqnarray}
Let us now define the function $F(\varthetaup)$ as follows
\begin{equation}
 F=\frac{d\textswab{g}}{d\varthetaup}
\end{equation}
Then the expression for the angular momentum reduces to:
\begin{eqnarray}
 J^{(\varrhoup)}_{\Delta}&=&\frac{-r_{\Delta}^{2}}{8\pi G}\oint_{S_{\Delta}}F
d\varthetaup\wedge\omega =
\frac{-r_{\Delta}^{2}}{8\pi
G}\oint_{S_{\Delta}}\textswab{g} f(\phi) d\omega
\end{eqnarray}
Now, note that we have previously defined the curvature of the
part of the connection $d\omega=2 (\rm{Im}\Psi_{2}\epsilon)$ where $\rm{Im}
\Psi_{2}$
provides us the information of the rotation.
Thus, if the solution is such that the Weyl tensor gives a contribution to
$\rm{Im}\Psi_{2} $, then the horizon admits an angular momentum. For spherical
symmetric solutions however,$\rm{Im}\Psi_{2} =0$. Thus for the case
of spherical solutions, the angular momentum is zero. Also note that
$\rm{Im}\Psi_{2} $ is gauge invariant and thus so is $J^{(\varrhoup)}_{\Delta}$.
This is interesting since the result is independent of the rescaling freedom of
$[\xi\ell^a]$,
the notion of angular momentum also makes sense for any NEH. It is important to
note that if we have any arbitrary vector field tangent to $S_{\Delta}$, one can
still define a Hamiltonian
$J_{\Delta}$ which will generate diffeomorphisms along that vector field.
However, we will 
not be able to identify that $J_{\Delta}$ to angular momentum as it is
intimately connected to
symmetries. We have obtained this result for a metric which has a symmetry
along some direction $\varrhoup^a$. This gave us the angular momentum for the
rotation abount the axis. In the case of spherical symmetry, the angular
momentum is zero because $\rm{Im}\Psi_{2} =0$. We also have two more Killing
vaectors. It is not very difficult to see that these two Killing vectors
give their corresponding angular momenta. However, again the momenta are
proportional to $\rm{Im}\Psi_{2} =0$ and hence are zero for spherical metric.
These results were previously derived in \cite{acs}. However, we rederive using
a choice of coordinate system for the sphere.

The first law for the rotating WIH can now be written down fairly easily
by taking into account the first laws for the two hamiltonian vector fields
considered above. Consider a vector field $W^{a}=\xi\ell^a -\Omega_{W}\varrhoup$
on the spacetime. The first law for this vector field will be:
\begin{equation}
\delta E_{\Delta}\=\frac{1}{8\pi G}\kappa_{(\xi\ell)}\delta (f
(\phi)\mathcal{A}_{\Delta})
+\Omega_{(W)}\delta J_{\Delta}
\end{equation}

%  Some comments are in order at this point . First, recall that the vector
%field
% $\varrhoup^a$
%  appearing in the formula for $J_\Delta$ is a fixed vector field on
%  $\Delta$. This was necessary for carrying out the Hamiltonian analysis.
%  Since $S$ is essentially an arbitrary cross-section of $\Delta$ , the fixed
% $\varrhoup^a$  need not
%  be tangent to $S$. However, the component of $\varrhoup^a$ tangent to defined
% by $\bar\varrhoup^a :=\varrhoup^a + (n.\varrhoup)\ell^a$
%  is a Killing vector of the two-metric $q_{ab}$ induced on $S$. We
%  could use $\bar\varrhoup^a$ to calculate $J_{\Delta}$ but if we had different
% foliation of $\Delta$ , then
%  we would have a different cross-section $S_0$ which would give a different
%  $\tilde\varrhoup^a$ \cite{ash5}. However, $J$ is independent of which
% $\varrhoup^a$ we use. Let
%  $S$ be given by $v = g(\varthetaup, \varrhoup)$ where $(v, \varthetaup,
% \varrhoup )$ are coordinates on $\Delta$ ; v is the affine parameter $\ell$.
% Then
% %
% \begin{equation}
% n = -dv + dg  ~~~~\ell^a =\elleft(\frac{\partial}{\partial v}\right)^a
% \end{equation}
% %
% Then, we have:
% %
% \begin{equation}
% \oint_{S_{\Delta}}(\bar\varrhoup.\omega)\epsilon-\oint_{S_{\Delta}}
%
%(\varrhoup.\omega)\epsilon\=\kappa_{(\ell)}\oint_{S_{\Delta}}(\ellie_{
%\bar\varrhoup
%}
% f)\=0
% \end{equation}
% %
% 
% 

\section{Spherical Horizon and Chern-Simons Symplectic
Structure}\label{sec:sph_hor}
Arguments of Bekenstein and Hawking, based on the laws of black hole
mechanics and semiclassical calculations, tells us that the entropy
of black holes are equal to quarter of their areas. However, such an
interpretation of entropy as area needs to be backed up by microstate 
counting \emph{a la} Boltzmann. Knowledge of microstates lies 
beyond the domain of classical theory because the laws of the
microscopic world are quantum mechanical. Thus, one needs a quantum theory of
spacetime for a satisfactory calculation of entropy.
One of the statistical interpretations of black
hole entropy is the {\em loop approach} based on Loop Quantum Gravity
\cite{abck, abk, ack}. Here, instead of knowing the 
microscopic degrees of freedom of the entire spacetime, it is proposed that
we consider their effects on the black hole horizon. The basic idea is
that the essential features of the black hole spacetime are captured by
some effective degrees of freedom on the horizon which originate 
because of the interaction of bulk and the boundary of
the spacetime. Isolated horizon formulation is relevant because such surfaces
capture the essential features of a black hole spacetime. One
determines the effective theory induced at an
isolated horizon, quantize it and count the appropriate quantum states. This
turn out to be consistent with the semiclassical estimates made by
Bekenstein and Hawking. The effective theory on the
horizon can only be a theory of the topological kind, namely it must
be insensitive to the metric on the horizon. This is because the
horizon is a null surface and therefore cannot support a physical
particle. The above papers show, through a detailed canonical phase
space analysis, that the effective theory on the horizon is
Chern-Simons type, more precisely a $U(1)$ Chern-Simons theory.

The main objective of the section is to find out the symplectic
structure of the effective field
theory on a spherically symmetric WIH of a fixed area, starting from the non-
minimally coupled Holst action in a completely covariant framework (the
derivation is similar in spirit to that in \cite{cg_holst}). We shall
show that the claims in the abovementioned papers are reinforced independent of
any slicing. Since the horizon is spherically symmetric, the boundary
conditions imply that the scalar field is constant
\emph{on the horizon}. Once these conditions are
fulfilled, it follows from the Einstein's equations that: 
\begin{equation}\label{sphere_phie_exp}
\Phi_{11}+\frac{1}{8}R -\frac{1}{2}\Lambda\=4\pi G e
\end{equation} 
where, $\Phi_{11} =\frac{1}{4}R_{ab}(\ell^{a}n^{b} + m^{a}\bar{m}^{b})$ and $e$
is spherically symmetric having contribution from the scalar field
the other minimally coupled fields like Maxwell fields \emph{etc}.
The equation (\ref{sphere_phie_exp}) implies that for spacetimes
with cosmological constant zero, the term $\Phi_{11}+\frac{1}{8}R$ is
spherically symmetric. Further, it is not diffcult to check that (see
\cite{cg_holst} for details)
\begin{equation}\label{sphere_grad_n}
\mathbf{\Psi_2} +\frac{1}{12}R\=\ellie_{\ell}~\mu +\kappa_{(\ell)}\mu
\end{equation}
where $\mu, \kappa_{(\ell)}$ are as previously defined (see
\ref{definitions_NP}).
All the terms on the right hand side of (\ref{sphere_grad_n})are real and $R$ is
real, the term $\mathrm{Im}\mathbf{\Psi_2}=0$,
\emph{i.e.}, $d\omega^{(\ell)}=0$. Then, the term $(~ \mathrm{Re}\mathbf{\Psi_2}
+\frac{1}{12}R~)$ is spherically symmetric.

Using the
equations (\ref{sphere_grad_n}) and (\ref{sphere_phie_exp}), we see
that the term $\mathcal{F}:=({\mathbf{\Psi_{2}}}^{(H)} -\Phi_{11} -
\frac{R}{24})$  is again 
spherically symmetric. Moreover, $\mathcal{F}$ is constant over $\Delta$. 
We want to find a value for this constant. It turns out that (see
appendix of \cite{cg_holst} for detailed calculations)
\begin{equation}\label{exp-dvh}
dV^{(m)}_{s}=-2i\mathcal{F} {}^2\mbf{\epsilon}
\end{equation}
The connection $iV^{(m)}$ is precisely the connection on the sphere $S^2$. Using
Gauss- Bonnet theorem, we get 
\begin{equation}\label{exp-psi-2}
({\mathbf{\Psi_{2}}}^{(H)} -\Phi_{11} -
\frac{R}{24})=-\frac{2\pi}{\mathcal{A}^{s}}
\end{equation}

Now, we define the connection for
$V^{(H)}=-im_{[I}\bm_{J]}A^{(H)}{}^{IJ}=(iV^{(m)}+\gamma\omega^{(\ell)})/2$.
Clearly, $dV^{(H)}=idV^{(m)}/2$ since on spherically symmetric horizon
$d\omega^{(\ell)}\=0$.
It follows from (\ref{exp-psi-2})that
\begin{equation}\label{epsilon_exp_sphere}
{}^2\mbf{\epsilon}=-\frac{\mathcal{A}^{s}_{\Delta}}{2\pi}~dV^{(H)}_{s}
\end{equation} 

We shall use this expression \ref{epsilon_exp_sphere} for the
surface contribution to symplectic current for fixed areas phase space. We shall
see that the surface symplectic structure is a $U(1)$ Chern-
Simons theory symplectic structure .

In (\ref{exp_j_delta_d}), the potential
$\psi_{(\ell)}$ is a function of $v$
only while $\mu_{(m)}$ is still a function of $(v, \varthetaup, \varrhoup)$.
A simple calculation gives
\begin{equation}
\int_{\Delta}J(\delta_{1}, \delta_{2})=\frac{1}{8\pi G \gamma}(\int_{S_{-}}-
\int_{S_{+}})\{ \delta_{1}\mu_{(m)}~ \delta_{2}{}^2\mbf{\epsilon}
-(1 \leftrightarrow 2)\}
\end{equation}

Now recall that the one-form $m$ has a rescaling freedom given by
$m\rightarrow g m(v,\theta,\varrhoup)=e^{-i\mu_{(m)} (v,\theta,\varrhoup)}
m(0,\theta, \varrhoup)$. This gives (see (\ref{trans_rule_uuv}) for the
transformation rules)
\begin{eqnarray}\label{exp_HolstV_sphere}
&&V_{s}^{(m)}\rightarrow V_{s}^{(m)}{}^{g}=V_{s}^{(m)}- i~d\mu_{(m)}\nonumber \\
&&V_{s}^{(H)}\rightarrow V_{s}^{(H)}{}^{g}= V_{s}^{(H)} +
\frac{1}{2}d\mu_{(m)}(v)
\end{eqnarray}
To proceed further, we use the expression (\ref{epsilon_exp_sphere}) in the 
symplectic current and integrate by parts and again use
(\ref{exp_HolstV_sphere}). This gives the following expression for the 
current
\begin{eqnarray}
\int_{\Delta} J(\delta_{1}, \delta_{2})&=& \frac{2f(\phi _s)}{8\pi G
\gamma}\frac{\mathcal{A}^{s}_{\Delta}}{\pi}(\int_{S_{-}}- \int_{S_{+}})\{
\delta_{1}V^{(H)}{}^{g}~\wedge \delta_{2}V^{(H)}{}^{g}- (1 \leftrightarrow 2)
\}\\ \nonumber
& -&\frac{f(\phi)}{8\pi G
\gamma}\frac{\mathcal{A}^{s}_{\Delta}}{\pi}(\int_{S_{-}}-
\int_{S_{+}})\{ \delta_{1}V^{(H)}{}^{g}\wedge\delta_{2}V^{(H)}- (1
\leftrightarrow 2)\}
\end{eqnarray}
Now, note that $ V^{(H)}{}^{g}$ is a function of $(v, \theta, \varrhoup)$
whereas $V^{(H)}$ has only the dependence on $(\theta, \varrhoup)$ 
and $\mu_{(m)}$ is  $v$ dependent. In other words, the $v$ dependence of 
$ V^{(H)}{}^{g}$ has been transferred to $\mu_{(m)}$ leaving $V^{(H)}$ only with
the angular dependence. Using this information, we get
\begin{equation}
\int_{\Delta} J(\delta_{1}, \delta_{2}) =\frac{f(\phi _s)}{8\pi G
\gamma}\frac{\mathcal{A}^{s}_{\Delta}}{\pi}(\int_{S_{-}}- \int_{S_{+}})\{
\delta_{1}V^{(H)}{}^{g}\wedge \delta_{2}V^{(H)}{}^{g}\}
\end{equation}
which is identical to the  Chern-Simons symplectic structure. The full
symplectic structure for the spherically symmetric and fixed phase space 
admitting a WIH
\begin{equation}
\Omega(\delta_{1}, \delta_{2})=\frac{1}{8\pi G\gamma}\int_{M}\left[
\delta_{1}(e^{I}\wedge
e^{J})~\wedge\delta_{2}A^{(H)}_{IJ} -\delta_{2}(e^{I}\wedge
e^{J})~\wedge\delta_{1}A^{(H)}_{IJ} \right] -\frac{f(\phi _s)}{8\pi G
\gamma}\frac{\mathcal{A}^{s}_{\Delta}}{\pi}\int_{S}\{
\delta_{1}V^{(H)}{}^{g}\wedge \delta_{2}V^{(H)}{}^{g}\}
\end{equation}
The boundary symplectic structure can be identified with that of a $U(1)$
Chern-Simons theory (because the symplectic structure involves only one type of
connection $V^{(H)}{}^{g}$) with level
$k={f(\phi_s)\mathcal{A}^{s}_{\Delta}}/4\pi G\gamma$. Upon quantization the
level becomes an integer. So ${f(\phi_s)\mathcal{A}^{s}_{\Delta}}/4\pi G\gamma$
has to be an integer in the quantum theory. This is a highly nontrivial result
of the WIH formlation. WIH phase space includes both extremal as well as
non-extremal types of horizons and what we find here is that the effective
theory on the boundary is insensitive to these two types of horizons. The Chern-
Simons gauge field does not see the $\xi$ scaling of the null normal $\ell^a$,
which controls the value of surface gravity
for the horizon. In the quantum theory one essentially counts the surface states
of the quantum Chern-Simons theory and hence, the entropy function is also
expected to be insensitive to two types of horizons.

\section{Discussions}
In this paper, we have studied the classical phase space of non-minimally
coupled Host action admitting a WIH. The boundary conditions for the WIH 
amiable for the non-minimal scalar coupling is stronger than the boundary
conditions in \cite{cg_holst, cg}. The non-minimally coupled
scalar field is constrained so that it remains constant along any $\ell^a$ in
the class $[\xi\ell^a]$. In the standard case of minimal couplings, the
constancy of the field along $\ell^a$ arises because of the energy conditions.
Non- minimally coupled scalar fields violate this energy conditions. Thus, this
condition is put in by hand and suffices for our discussion. The zeroth law
for the WIH is a result of these boundary conditions. Next,
we have constructed the symplectic structure on this phase space and proved
the first law. The essence of this derivation is similar to that in
\cite{cg_holst, cg}. We also included rotations and showed in a
detailed derivation how the first law is changed. We have not
included any minimally coupled fields like the Maxwell field. The
result of such inclusion is standard and can be derived as in \cite{cg_holst,
cg}. Subsequently, we have
shown that on a fixed area phase space containing spherical horizons, the
symplectic structure is that of a $U(1)$ Chern-Simons theory. This result 
was only derived for the constant class of null-normals  $[c\ell^a]$ (Weakly
Isolated Horizons) in a covariant formalism (see \cite{ac}). We rederive these
results for WIH in a covariant phase space formulation. It is easy to map
the covariant phase space results to canonical phase space (see appendix
of\cite{cg_holst}).

Let us point out the importance of these results. It is known that string
theories generally predict non-minimal scalar field couplings. Moreover,
extremal black holes are ubiquitous in such theories \cite{mohaupt}. The
standard prescription for calculating the entropy of these extremal black
holes is the Wald's Noether charge approach \cite{Wald}. It is well known that 
this formulation is not well
suited for the extremal black holes. Indeed, in this case the extremal and non-
extremal black holes belong to distinct phase spaces (see \cite{cg_holst} for a
detailed discussion). Thus, it is difficult to make sense of extremal limits
of quantities defined for non- extremal black holes. On the other hand, WIH
naturally encompass both extremal and non-extremal black holes horizons. The
space of solutions admitting WIH internal boundary
has both the extremal and non- extremal solutions on the same footing.
On this phase space, extremal black holes can be defined as a limit of sequence
of non-extremal black holes. The laws of black hole mechanics for these two
type of horizons are clarified in this formulation. 
We have also showed that the topological theory induced on the horizon is a
$U(1)$ Chern- Simons theory 
induced on WIH is irrespective of the nature of extremality of the solution.
Indeed the Chern-Simons gauge field $V^{(H)}$ does not depend on the $\xi$
function which controls the extremality/non-extremality of the horizon. 
It will be argued thus that if the space of solutions has both extremal and non-
extremal horizons, the 
entropy is same for both species. This is because, one essentially counts the
surface states of the quantum Chern-Simons theory and hence, the entropy
function is also expected to be insensitive to two types of horizons. In
other words, WIH is ideally suited for study of extremal and non-extremal
black holes in the same footing. Thus, the formalism is ideally suited for
study of black holes in string theory.

\section{Acknowledgments}

The author thanks Prof. A. Ghosh for discussions on various aspects of the
work. The author also acknowledges discussions with Prof. P. Majumdar.

\end{document}